\documentclass[twocolumn, prd, nofootinbib,
showpacs, preprintnumbers, amsmath, amssymb]
              {revtex4}

\usepackage{graphicx}

\begin{document}

\title{ Dark energy due to effective quantum field theory }

\author{ Michael~Maziashvili}
\email{maziashvili@iphac.ge} \affiliation{ Andronikashvili
Institute of Physics, 6 Tamarashvili St., Tbilisi 0177, Georgia }

\begin{abstract}

In the cosmological context an effective quantum field theory describing
the behavior of visible matter in the universe is characterized with its
inherent UV cutoff and also with an IR scale that is set by the
cosmological (particle) horizon. This UV - IR relation naturally
defines a space-time grid over a horizon scale. Using the
approach for determining of dark energy through the space-time uncertainty
relation versus such a space-time grid, we estimate the energy density
and pressure of a dark energy defined by this UV - IR relation. Such a
dark energy shows up to decay linearly with time and exhibits a
negative pressure only recently.

\end{abstract}


%   04.60.-m - Quantum gravity

%   06.20.Dk - Measurement and error theory

%   98.80.-k  Cosmology

%   95.36.+x    Dark energy

\maketitle

\section*{Introduction}

Having so successful theories of ordinary matter after the study
of all these centuries, in $1998$ astronomers informed us that
ordinary matter constitutes only about $5$ percent of the whole
mass of the universe \cite{darkenergy}. About $70$ percent of the
mass of the universe is made of what we call now dark energy and
about $25$ percent of dark matter. Hitherto very little is known
about the dark matter, and even less about the dark energy. What
we see are their global gravitational effects. They neither emit
nor absorb light to any significant extent and generically they
seem to interact very feebly not only with photons, but with
ordinary matter altogether. Dark energy, because of which the
expansion of the universe has recently begun to accelerate, is
equally dense everywhere (as far as we can tell) as if it is an
intrinsic property of space-time itself. It was noticed long ago
by Zeldovich that one of the possible sources for dark energy
might be QFT vacuum energy \cite{Zeldovich}. Namely, as the QFT
respects Lorentz invariance, the vacuum energy mimics the
cosmological constant, to wit \begin{equation}\label{Lorentzinv}
\langle 0|T_{\mu\nu}|0\rangle = \langle 0|T_{00}|0\rangle
\,\eta_{\mu\nu}~,\end{equation} where $\eta_{\mu\nu}$ is a Lorentz
metric $\eta_{00} = -\eta_{11} = -\eta_{22} = -\eta_{33} = 1$.
Unfortunately vacuum energy density defined as a {\tt
Nullpunktsenergie} appears to be infinite. However, the infinity
arises from the contribution of modes with very small wavelengths
and for we do not know what actually might happen at such scales
it is reasonable to introduce a cutoff and hope that a more
complete theory will eventually provide a physical justification
for doing so. But this is not all the story, as in QFT the
energy-momentum operator $T_{\mu\nu}$ (and correspondingly the
source of gravity $\langle 0|T_{\mu\nu}|0\rangle $) is not
uniquely defined because of operator ordering. In the framework of
QFT we are usually subtracting this (divergent) vacuum energy
which is equivalent to the normal operator ordering in
$T_{\mu\nu}$. Or equivalently in the path integral approach one
observes that the equations of motion for matter fields are
invariant under the shift of the matter Lagrangian by a constant
that results in a new energy momentum tensor
\[T_{\mu\nu}~~\rightarrow ~~T_{\mu\nu} +
\mbox{const.}\,\eta_{\mu\nu} ~.\] Thus we need some physical
principle that could guide us into this problem to guess what is
the interplay between QFT and dark energy. Motivated by the
definition of dark energy through the space-time uncertainty
relation (STU) \cite{Sasakura}, we consider similar way for
defining the vacuum energy density for QFT describing the behavior
of matter at different stages of the cosmological evolution.

\subsection*{Energy budget of space-time due to STU}

STU for a given background space implies a
principal limitation on the space-time distance measurement in light of quantum
mechanics and general relativity. The physical
meaning of this sort of relation is that during the measurement of
some length scale we are disturbing the background space-time, and in
view of quantum mechanics and general relativity it turns out that
this disturbance can not be reduced arbitrarily. (Throughout this
paper we assume natural system of units $\hbar = c = 1$). Taking the
Minkowskian background space, on the quite general grounds we notice
that a measuring device (or simply a test body) with zero mean
velocity, having the mass $m$ and located within the region $\delta x$,
is characterized by the gravitating energy \begin{equation} \label{gravdisturbance} E = m + {\delta p^2 \over
m} ~,\end{equation} where $\delta p \simeq \delta x^{-1}$. The second term in this
equation accounts for the quantum fluctuation energy of a measuring
device. If we are interested in measurement of some local
characteristics of the background space, $\delta x$ can not be taken
arbitrarily large. Therefore minimizing the
Eq.(\ref{gravdisturbance}) with respect to $m$ one gets an
unavoidable gravitational disturbance of the background
space-time. Combining the Heisenberg uncertainty relations with
general relativity, K\'arolyh\'azy obtained STU for Minkowski space of the form \cite{Karolyhazy}
\begin{equation}\label{Karol} \delta t \gtrsim
  t_P^{2/3}t^{1/3}~,\end{equation} which tells us that one can not
measure the space-time distance $t$ in Minkowski space to a better accuracy than
$\delta t$. This relation was studied further
from different points of view in \cite{NvD, Sasakura}. STU naturally
translates into the metric fluctuations, for if it was possible to
measure the background metric precisely one could estimate the length between
two points exactly. As we are dealing with the Minkowskian
background space the rate of metric fluctuations over a length
scale $t$ can be simply estimated as $\delta g_{\mu\nu} \sim
\delta t / t$. We naturally expect there to be some energy density
associated with these fluctuations. One can use the following
simple reasoning for estimating the energy budget of Minkowski
space \cite{Sasakura}. With respect to the STU relation a length scale
$t$ can be known maximum with a precision $\delta t$ determining thereby a
minimal detectable cell $\delta t^3$ over a spatial region $t^3$.
Such a cell represents a minimal detectable unit of space-time
over a region $t^3$ and if it has a finite age, $t$, its existence
due to time energy uncertainty relation can not be justified with
energy smaller then $t^{-1}$. Hence, having the STU relation one
concludes that if the age of Minkowski space is $t$ then over a
spatial region with linear size $t$ (determining the maximal
observable patch) there exists a minimal cell $\delta t^3$ the
energy of which due to time-energy uncertainty relation can not be
smaller than $t^{-1}$ leading to

\[\rho_{STU} \sim {1 \over t \delta t^3}~.\] Using the
Eq.(\ref{Karol}) one gets \begin{equation}\label{studarken}
\rho_{STU} \sim {1 \over t_P^2  t^2}~,\end{equation} which for
$t_0 \sim H_0^{-1}$ gives pretty good value for the present dark
energy density. Recently such a parametrization of dark energy by
the age of the universe was studied in much details
\cite{Agegraphic}. Two major problems one may expect in such
models are as follows. The radiation and matter behave also as
$\sim t^{-2}$, correspondingly the present coincidence of the dark
energy density with the matter density makes it obscure why such
dark energy should become dominant, for instance one may expect
its pressure like the matter to be zero \cite{Hsu}, and for the
same reason it may be hard to reconcile such dark energy models
with the early cosmology \cite{PadBarr}. So on the quite general
grounds one finds that it may be troublesome to avoid these
problems in more or less natural way. Nevertheless, it should be
noticed that the parametrization of this kind of dark energy by
the conformal time allows one to avoid those problems
\cite{Newagegraphic}.

\subsection*{STU versus the UV - IR relations in QFT}

Imposing gravitational bounds on an effective quantum field theory one arrives
at the relations between UV and IR scales
\cite{CKN}, which on the other hand can be viewed as a space-time
uncertainty relations coming from a space-time measurement
\cite{mazia}.

For an effective quantum field theory in a box of size
$l$ with UV cutoff $\Lambda$ the entropy $S$ scales as, \[ S \sim
l^3\Lambda^3~.\] That is, the effective quantum field theory counts the
degrees of freedom simply as the numbers of cells $\Lambda^{-3}$ in
the box $l^3$. Nevertheless, considerations involving black holes demonstrate
that the
maximum entropy in a box of volume $l^3$ grows only as the
area of the box \cite{Bekenstein} \[S_{BH} \simeq \left({l \over l_P}\right)^2~.\]
So that, with respect to the Bekenstein bound \cite{Bekenstein} the
degrees of freedom in the volume should be counted by
the number of surface cells $l^2_P$. A consistent physical picture can be constructed
by imposing a relationship between
UV and IR cutoffs \cite{CKN}
\begin{equation}\label{BHbound}l^3 \Lambda^3 \lesssim S_{BH} \simeq
\left({l\over l_P}\right)^2~,
\end{equation}
where $S_{BH}$ is the entropy of a black hole of size $l$.
Consequently one arrives at the conclusion that the length $l$,
which serves as an IR cutoff, cannot be chosen
independently of the UV cutoff, and scales as $\Lambda^{-3}$.
Rewriting
this
relation wholly in length terms, $\delta l \equiv \Lambda^{-1}$,
one arrives at the Eq.(\ref{Karol}).

Another space-time uncertainty relation is based on the
random walk approach to the distance measurement, see \cite{Amelino}
and the last paper in \cite{mazia}. Gravitational field is described
in terms of space-time metric, so figuratively speaking it measures
space-time distances. To measure the space-time distance gravitational
field has the only intrinsic length scale $l_P$. If we assume our
ruler is just $l_P$, that is, both its length and precision are given
by the Planck length, we arrive at the equation

\begin{equation}\label{rand}\delta l \gtrsim
  \l_P\left({l\over l_P} \right)^{1/2} = \,
  l_P^{1/2}l^{1/2}~.\end{equation} An effective quantum field theory
  has its own explanation of this relation. In effective
quantum field theory the energy density of the vacuum is set by the UV
  cutoff as $\sim \Lambda^4$. The
gravitational radius associated with the vacuum energy of the
system, $E_{vacuum}\sim l^3\Lambda^4 ~\Rightarrow ~r_g\sim
l_P^2l^3\Lambda^4$, will be greater than the size of the system,
$l$, if UV cutoff is defined from the Eq.(\ref{BHbound}). To be on
the safe side, one can impose stronger constraint requiring the
size of the system to be greater than the gravitational radius
associated to the maximum energy of the system \cite{CKN}
\begin{equation}
\label{blueberunium} l_P^2l^3 \Lambda^4 \lesssim l \, .
\end{equation} With respect to this relation the IR cutoff scales like $\Lambda^{-2}$. This relation
written in
 length
terms ($\delta l \equiv \Lambda ^{-1}$) is the
 Eq.(\ref{rand}).

So, we see the effective quantum field theory picture
behind the Eqs.(\ref{Karol},\,\ref{rand}). The
space-time uncertainty relations,
Eqs.(\ref{Karol},\,\ref{rand}), in their turn shed
new light on the above UV - IR relations
(\ref{BHbound},\,\ref{blueberunium}) obtained in the framework of
effective field theory in \cite{CKN}, exhibiting that the IR scale
can not be known to a better accuracy than $\delta l$ representing thereby a
lower bound on the admissible UV scale (in length terms).

The main point we want to draw from this section for what follows is
that an effective quantum filed theory characterized with UV and IR
scales defines a space-time grid that can be considered versus the space-time uncertainty relation.

\subsection*{Defining dark energy due to QFT}

In most QFT applications with an UV cutoff it is customary to set the vacuum energy density (simply on the
dimensional grounds) as $\sim \Lambda ^4$. This energy density is
understood to come from {\tt Nullpunktsenergie}. In this regard two
remarks are in order. First and main remark as it was discussed in the
introduction is that usually in the framework of QFT the vacuum energy $H|0\rangle =
E_0|0\rangle$ is treated as an unphysical quantity that may be set
arbitrarily\footnote{For a crystal the {\tt Nullpunktsenergie}
represents the vibration energy of crystal molecules at a zero
temperature. This energy manifests itself even at a finite
temperature, and has therefore quite definite physical meaning. One can see for instance very readable popular book by Kaganov
\cite{Kaganov}, where many conceptual points of condensed matter
physics are elucidated.}. The second remark is more technical and has to do with the
regularization procedure. Namely estimating the {\tt Nullpunktsenergie} of QFT in the Minkowskian
background, we should care the Eq.(\ref{Lorentzinv}) to be satisfied
\cite{eqofstate}. Regularizations of the {\tt Nullpunktsenergie} which
respect the Lorentz symmetry of the underling theory disfavor its
quartic dependence on the UV scale, but rather it appears to depend
quadratically on the UV scale \cite{eqofstate}. This point has attracted little
attention hitherto for many authors still follow the old
customary. One may object that the presence of a IR scale immediately
brakes the Lorentz invariance and through the box boundary conditions
naturally leads to this estimate of {\tt Nullpunktsenergie} (as we did
in the previous section). But we should recall that in the
cosmological context IR scale set by the particle horizon defines
merely a causally connected region and does not imply any box boundary
conditions at this scale. In what follows we will skip this {\tt
  Nullpunktsenergie} paradigm.

The behavior of matter in the universe at different stages of its evolution is described by the particle
physics models, which in the framework of an effective quantum field
theory are characterized with their intrinsic UV energy scales \cite{Rubakov}. The
microscopic energy scales of quantum mechanics and the macroscopic
properties of our present Universe are intimately connected. For
instance, the $O$(eV) energy scale of atomic physics manifests itself
through the existence of the cosmic microwave background radiation,
and the $O$(MeV) scale of nuclear physics through the primordial
origin of light element abundances. The connection can be extended
further. As we move further back in time, several phase transitions in
the universe might be available. One can order the
sequence of early time cosmological phase transitions roughly as: The
GUT phase transition when the universe was about $\sim 10^{-35}$s old and
the temperature about $T_{GUT} \sim 10^{16}$\,GeV\,;\footnote{Let us
  notice that many models
of inflation indicate that the universe never had such a high
temperature after inflation.} the EW phase transition
when the universe was about $\sim 10^{-12}$s old with a temperature $T_{EW}
\sim 100$\,GeV\,; the QCD phase transition at about $\sim 10^{-5}$s when the
temperature was about $T_{QCD} \sim 170$\,MeV. So that the UV scale suggested by the particle physics
describing the behavior of matter in the universe at different stages does not follow
neither Eq.(\ref{Karol}) nor Eq.(\ref{rand}) (or equivalently neither
Eq.(\ref{BHbound}) nor Eq.(\ref{blueberunium})). Taking the $\Lambda(t)$ that follows from particle
physics models describing different stages of the universe and straightforwardly
repeating the discussion of the second section we get\footnote{Motivated with Dirac's large number hypothesis similar
  expression was suggested by Zee in \cite{Zee2}. }   \begin{equation}\label{qftdarken} \rho_{QFT} \simeq {\Lambda^3(t)
  \over t}~. \end{equation} From above discussion one infers that after
the nucleosynthesis (which started when the universe was about $\sim
1$s old) we can take $\Lambda = O$(MeV). Taking $\Lambda$ to be about
$\Lambda \simeq 100$\,MeV after the nucleosynthesis, from Eq.(\ref{qftdarken}) one
gets pretty good value for the present dark energy density. As this energy density decays
linearly it will not affect the early time cosmology. The equation of
state can be simply estimated. Assuming that this energy component
dominates \[ H^2 = {8\pi \over 3m_P^2} \rho_{QFT}~,  \] and using
energy-momentum conservation \[ p = -{\dot{\rho} \over 3 H} - \rho~,
\] one gets \begin{equation}\label{pressure}  p_{QFT} \simeq \sqrt{ m_P^2\Lambda^3\over 24\pi t^3 } -
   {\Lambda^3 \over t}~. \end{equation} The second term in Eq.(\ref{pressure}) becomes dominant for \begin{equation}\label{domination} t \gtrsim {m_P^2
   \over 24\pi \Lambda^3 } \simeq
   10^{58}t_P~. \end{equation} So this dark energy exhibits a negative
   pressure just recently.

\section*{Concluding remarks}

In the cosmological context we are operating with two length scales, the
IR and UV ones, where IR scale is naturally set by the horizon while
UV scale is determined by the effective quantum field theory
describing the behavior of matter in the universe. These two length scales uniquely define a space-time grid over a
causally connected region, or simply a causally connected space-time
grid that can be studied versus the space-time uncertainty
relation. Namely, assuming the finiteness of the age of a space-time,
$t$, we conclude that due to time-energy uncertainty relation the spatial
cell $\Lambda^{-3}$ set by the UV scale over the observable region can
not be smaller than $t^{-1}$, that immediately leads to the
Eq.(\ref{qftdarken}). After the QCD phase transition in the universe
we can take the UV scale to be in the $O$(MeV) range. Taking $\Lambda
\simeq 100$MeV after the quark confinement transition in the universe,
one gets pretty good value for present dark energy density and also
that value of the UV scale guaranties negative pressure at the present
stage, Eqs.(\ref{pressure},\,\ref{domination}), as befits a dark energy. The advantage of the QFT dark energy model,
Eq.(\ref{qftdarken}), over the STU dark energy, Eq.(\ref{studarken}), is
that as it decays linearly with time it can not spoil the successes of early
cosmology and, on the other hand, it obviously exhibits a negative pressure for
the present cosmological stage. As an interesting feature the pressure
of this dark energy, Eq.(\ref{pressure}), becomes negative only
recently, Eq.(\ref{domination}). So that the early time cosmology is
doubly protected from the action of this dark energy. Certainly the
validity of this sort of dark energy requires further detailed analysis
against the observed cosmological data.

Author is greatly indebted to Professor Alexander Vilenkin for
invitation and hospitality at the \emph{Tufts Institute of
Cosmology}, where this paper was done. Author is also indebted to
Professors Rong-Gen Cai and Naoki Sasakura for stimulating
comments. This work was supported by the \emph{CRDF/GRDF}, the
\emph{INTAS Fellowship for Young Scientists} and the
\emph{Georgian President Fellowship for Young Scientists}.

\end{document}